\RequirePackage{lineno}
\documentclass[aps,prb,floatfix,british]{revtex4}

\usepackage{amssymb}
\usepackage{setspace}
\usepackage{amsmath}
\usepackage{mathrsfs}
\usepackage{amssymb}
\usepackage{babel}
\usepackage{graphicx}
\usepackage{epstopdf}
\usepackage[sort&compress]{natbib}
\usepackage{topcapt}
\usepackage{textcomp}
\usepackage{xcolor}
\usepackage[colorlinks,urlcolor=blue,linkcolor=blue,citecolor=blue]{hyperref}

\begin{document}
\begin{spacing}{2.0}

\title{Global Correlation and Local Information Flows \\
in Controllable Non-Markovian Open Quantum Dynamics}

\author{Xin-Yu Chen,$^{1}$
Na-Na Zhang,$^{2,3}$
Wan-Ting He,$^{2}$\\
Xiang-Yu Kong,$^{1}$
Ming-Jie Tao,$^{4}$
Fu-Guo Deng,$^{2}$
Qing Ai,$^{2,}$\footnote{aiqing@bnu.edu.cn}
Gui-Lu Long$^{1,}$ \footnote{gllong@tsinghua.edu.cn}
}

\address{$^{1}$Department of Physics, Tsinghua University, Beijing 100084, China \\
$^{2}$Department of Physics, Applied Optics Beijing Area Major Laboratory,
Beijing Normal University, Beijing 100875, China \\
$^{3}$School of Optoelectronics Engineering, Chongqing University of Posts and Telecommunications, Chongqing 400065, China \\
$^{4}$Faculty of Fundamental Science, Space Engineering University, Beijing
101400, China}

%
%
%
%
%
%
%

\date{\today}

\begin{abstract}
In a fully-controllable experiment platform for studying non-Markovian open quantum dynamics, we show that the non-Markovianity could be investigated from the global and local aspects. By mixing random unitary dynamics, we demonstrate non-Markovian and Markovian open quantum dynamics. From the global point of view, by tuning the base frequency we demonstrate the transition from the Markovianity to the non-Markovianity as measured by the quantum mutual information (QMI). In a Markovian open quantum process, the QMI decays monotonically, while it may rise temporarily in a non-Markovian process. However, under some circumstances, it is not sufficient to globally investigate the non-Markovianity of the open quantum dynamics. As an essential supplement,
we further utilize the quantum Fisher information (QFI) flow to locally characterize the non-Markovianity in different channels. We demonstrate that the QMI in combination with the QFI flow are capable of measuring the non-Markovianity for a multi-channel open quantum dynamics.


\end{abstract}

\maketitle


\section*{Introduction}

Quantum coherence and entanglement lie at the heart of important resources for quantum metrology and quantum information processing \cite{Degen2017,Feng2013}.
Due to interaction with the environment,
any quantum system inevitably suffers from decoherence and quantum
entanglement disappears at a finite time \cite{Yu2004,Almeida2007}. However, environment does not always play a harmful role and some interesting phenomena may rise in open quantum dynamics with structured baths. For example, in a spin bath, multi-decoherence processes may be slower than single-decoherence processes when applied with appropriate pulse sequences \cite{Zhao2011,Huang2011}. In radical-pair mechanism for avian compass, the local magnetic environments compete against the homogeneous geomagnetic field to affect the yield of chemical reaction and thus effectively play the role of measuring instrument for the weak geomagnetic field \cite{Cai2013,Cai2012}. Interestingly, the environment can assist efficient energy transfer towards the reaction center in natural photosynthesis \cite{Lambert2013,Ai2013}. In a non-Markovian environment, the quantum Zeno effect can be utilized to steer the state evolution and explore quantum entanglement to improve the precision of metrology \cite{Ai2013-2,Harrington2017,Ai2010,Chin2012}.

Generally speaking, the dynamics of open quantum systems are described by the quantum master equation, in which the well-known Markovian and Born approximations have been applied \citep{Breuer2007}.
However, under some circumstances, the quantum dynamics can significantly
deviate from that predicted by the Markovian quantum master equation,
e.g. strong coupling to some vibrational modes in the bath \citep{Chin2013},
or the system-bath couplings are comparable to the intra-system couplings \cite{Ishizaki2009-1}. Furthermore, the positivity may be violated in the reproduced quantum dynamics by the non-Markovian quantum master equation when some approximations are made in the deduction \cite{Piilo2008,Piilo2009,Breuer2009-2,Ai2014}. Therefore, it is necessary to exactly describe the open quantum dynamics for the whole parameter regime in a unified manner \cite{Tao2020}, e.g. the hierarchical equation of motion (HEOM) \cite{Ishizaki2009-2}.

Apart from the theories for simulation, many
useful measures have been put forward to assess the non-Markovianity
of the open quantum dynamics \citep{Breuer2016,Vega2017,Li2018,Haase2018,Wu2020}.
Because the Markovian
open quantum process will erase the memory of the initial state,
quantum systems which start from different initial states will result in
the thermal equilibrium. The temporary rise in the trace distance between
two initial states can be viewed as a signature of the non-Markovianity
\citep{Breuer2009-1,Liu2011}. Because local trace-conserving complete positive
maps do not raise the entanglement, in Markovian processes the monotonic
decay will be observed for the entanglement between the system and
an ancillary system. Thus, the entanglement was also proposed for
measuring the degree of non-Markovian behavior in open quantum dynamics \cite{Rivas2010}, e.g. in photosynthetic exciton energy transfer \cite{Sarovar2010,Ai2014}.
However, because the entanglement characterizes only the quantum correlation \citep{Groisman2005}, it may be more convincing to use the quantum mutual information (QMI), which is the total correlation including both quantum and classical correlation, to explore the non-Markovianity \citep{Luo2012}.
Nevertheless, all the above measures only quantify the non-Markovianity from a global point of view. Since a many-body system can interact with the environments individually, it is quite natural to ask can we investigate the non-Markovianity from a different point of view, e.g. from different channels. The flow of the quantum Fisher information (QFI), which has been used to estimate the error bound of quantum metrology, was proposed to study the information exchange between the system and bath \cite{Lu2010}. In Markovian processes, the QFI flow is unidirectional from the system to the bath, while it is bidirectional in non-Markovian processes. Especially, the QFI flows in different channels can be used to study the local information flows between the system and bath, which can not be explored with the global measures. On the other hand, it was suggested that mixing Markovian semigroups or unitary dynamics may result in the emergence of non-Markovianity by virtue of an ancillary system \cite{Breuer2018}. It might be interesting to experimentally engineer and mix unitary quantum dynamics to simulate a non-Markovian open quantum dynamics.

In order to experimentally investigate the non-Markovianity from both the global and local points of view, an experiment platform with both systematic and environmental parameters accurately tunable is required. Recently, we have theoretically developed and experimentally demonstrated an efficient quantum simulation approach, which can exactly simulate the open quantum dynamics for an arbitrary Hamiltonian and various types of spectral densities \cite{Wang2018,Zhang2020,Buluta2009,Georgescu2014}. It effectively makes use of the bath-engineering technique \cite{Soare2014} and the gradient ascent pulse engineering (GRAPE) algorithm \cite{Khaneja2005,Li2017} to exponentially accelerate the exact simulation.

\section*{Results}

\bigskip
\subsection*{Two-Qubit System}

In open quantum systems, their dynamics are governed by the total Hamiltonian \citep{Breuer2007}
\begin{eqnarray}
{H_{\mathrm{OQD}}}={H}_{\rm{S}}+{H}_{\rm{B}}+{H}_{\rm{SB}},
\end{eqnarray}
where the system Hamiltonian of multi-level $\vert i\rangle$'s with energy $\varepsilon_{i}$ and couplings $J_{ij}$'s is ${H}_{\rm{S}}=\sum_{i}\varepsilon_{i}\vert i\rangle\langle i\vert+\sum_{i\neq j}J_{ij}\vert i\rangle\langle j\vert$,
and the Hamiltonian of the bath reads ${H}_{\rm{B}}=\sum_{i,k}\omega_{k}a_{ik}^{\dagger}a_{ik}$ with $\omega_{k}$ and $a_{ik}^{\dagger}$ ($a_{ik}$) being frequency and creation (annihilation) operator of $\vert i\rangle$'s $k$th harmonic oscillator.
Generally, the system-bath Hamiltonian can be of arbitrary form \cite{Soare2014} and
without loss of generality we assume a pure-dephasing form ${H}_{\rm{SB}}=\sum_{i,k}g_{ik}\vert i\rangle\langle i\vert(a_{ik}^{\dagger}+a_{ik})$ with coupling $g_{ik}$.
All the information about the system-bath interaction is given by
the spectral density $G(\omega)=\sum_{i,k}g_{ik}^{2}\delta(\omega-\omega_{k})$ \cite{Breuer2007}.

\begin{figure}
\includegraphics[bb=0 0 420 180,width=8.7cm]{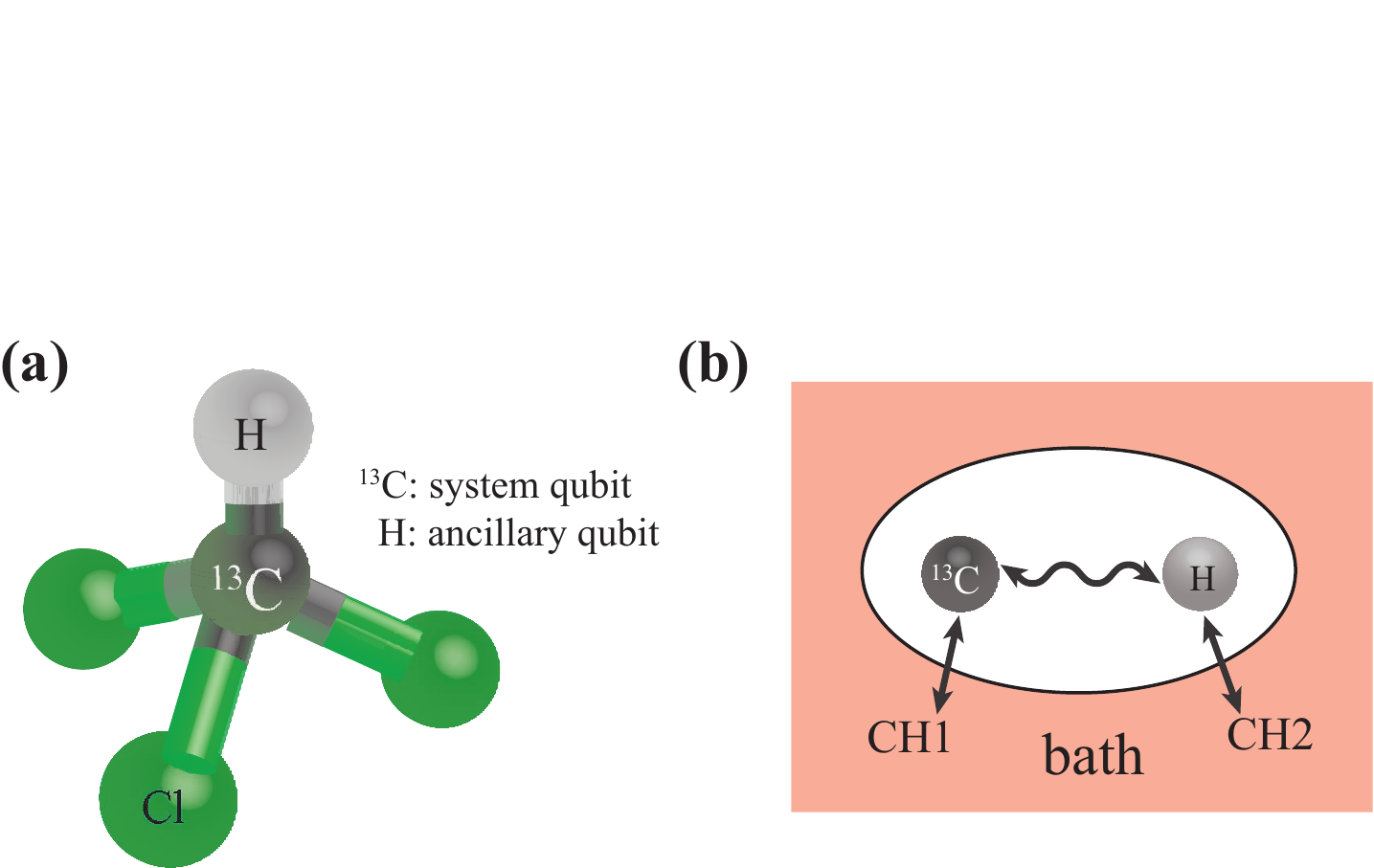}
\caption{Schematic for quantum simulation of non-Markovian open quantum dynamics. (a) $^{13}$C-labelled chloroform for the quantum simulation in NMR. (b)
Carbon atom acts at as the system qubit, while the hydrogen atom is the ancillary qubit. \label{fig:Scheme}}
\end{figure}

Here, we consider a two-qubit system, one of which is the system qubit and the other is the auxiliary qubit.
In Fig.~\ref{fig:Scheme}(a), we show the $^{13}$C-labelled chloroform for the quantum simulation, where $^{13}$C is the system qubit and H is the auxiliary qubit.
In the quantum simulation, the total Hamiltonian reads
\begin{eqnarray}
{H}={H}_{\rm{S}}+{H}_{\rm{N}},
\end{eqnarray}
where ${H}_{\rm{S}}$ and ${H}_{\rm{N}}$ correspond to the system Hamiltonian and the noise Hamiltonian, respectively.
The system Hamiltonian is
\begin{eqnarray}
{H}_{\rm{S}}=\frac{\omega_{\rm{s}}}{2}\sigma_{\rm{s}}^{z}+\frac{\omega_{\rm{a}}}{2}\sigma_{\rm{a}}^{z},
\end{eqnarray}
where the subscript ${\rm{s}}$ (${\rm{a}}$) labels the system (ancillary) qubit with $\omega_{i}$ and $\sigma_{i}^{z}$ being Zeeman energy and Pauli operator ($i={\rm{s,a}}$), respectively. Initially,
the total system is prepared in the maximum-entangled state
\begin{eqnarray}
\left|\psi_{\rm{sa}}\left(0\right)\right\rangle =\left(\left|00\right\rangle+\left|11\right\rangle\right)/\sqrt{2}.
\end{eqnarray}
We consider that only the system qubit evolves under the influence of the noise,
with the noise Hamiltonian
${H}_{\rm{N}}=\beta_{\rm{s}}(t)\sigma_{\rm{s}}^{z}$.
We can obtain the density matrix of the
total system $\rho_{\rm{{sa}}}\left(t\right):=\left(\Lambda_{t}\otimes I\right)\rho_{\rm{{sa}}}$ as
\begin{eqnarray}
\rho_{\rm{sa}}\left(t\right)=[\vert00\rangle\langle00\vert+\vert11\rangle\langle11\vert+g(t)\vert00\rangle\langle11\vert+g^{\ast}(t)\vert11\rangle\langle00\vert]/2,
\end{eqnarray}
where the off-diagonal term is
\begin{eqnarray}
g(t)=\exp[-i(\omega_{\rm{s}}+\omega_{\rm{a}})t]f(t)
\end{eqnarray}
with the norm $f(t)=\exp[-2\chi(t)]$ and the decoherence
factor $\chi\left(t\right)=\frac{4}{2\pi}\smallint_{-\infty}^{+\infty}\frac{d\omega}{\omega^{2}}S_{ii}\left(\omega\right)\sin^{2}(\frac{\omega t}{2})$.
Here the noise power spectrum $S_{ii}\left(\omega\right)=\smallint_{-\infty}^{+\infty}d\tau\left\langle \beta_{i}\left(t+\tau\right)\beta_{i}\left(t\right)\right\rangle e^{-i\omega\tau}$
is the Fourier transform of the noise correlation function
$\left\langle \beta_{i}\left(t+\tau\right)\beta_{i}\left(t\right)\right\rangle =\underset{T\rightarrow\infty}{\lim}\frac{1}{2T}\smallint_{-T}^{T}dt\beta_{i}\left(t+\tau\right)
\beta_{i}\left(t\right)=\left(\frac{\alpha}{2}\right)^{2}\sum_{j=1}^{J}\left[F\left(\omega_{j}\right)
\omega_{j}\right]^{2}\left(e^{i\omega_{j}\tau}+e^{-i\omega_{j}\tau}\right)$,
where the first equation is valid in the large-ensemble limit \citep{Goodman2015}.
Here, $\beta_i\left(t\right)$ represents the distribution of noise in time domain, and it can be written as
$\beta_{i}(t)=\alpha\sum_{j=1}^{J}F_{i}(j)\sin(\omega_{j}t+\psi^{(i)}_{j})$, where $\alpha$ is the noise amplitude, $\omega_{j}=j\omega_{0}$ with base frequency $\omega_{0}$ and cut-off frequency $J\omega_{0}$, $\psi^{(i)}_{j}$'s are random numbers.
And the types of noise rely on the function $F\left(\omega_{j}\right)$.

\bigskip
\subsection*{Quantum Mutual Information}

\begin{figure}
\includegraphics[bb=0 140 1200 940,width=8.7cm]{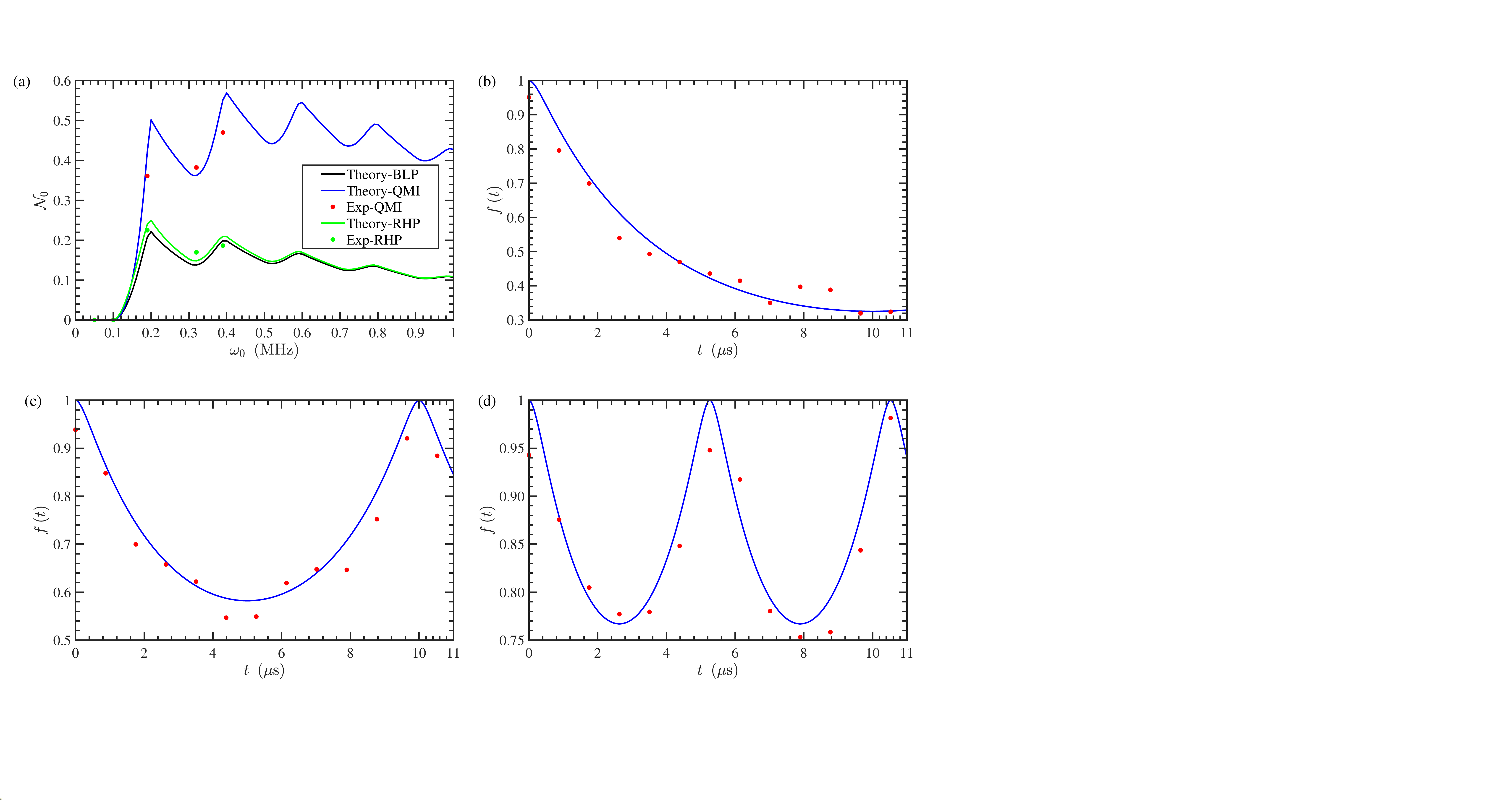}
\caption{Non-Markovianity measured by global correlation. (a) Phase diagram of QMI $\mathcal{N}_0$ vs base frequency
$\omega_{0}$. Quantum dynamics of off-diagonal term in the density
matrix $f(t)$ for (b) $\omega_{0}=0.05$~MHz; (c) $\omega_{0}=0.1$~MHz;
(d) $\omega_{0}=0.19$~MHz. Other parameters are $\gamma_\textrm{NMR}=0.9$~MHz, $\lambda_\textrm{NMR}=0.2$~kHz, $\omega_{J}=5$~GHz, $T_\textrm{OQD}=3\times10^4$~K, $N=150$. The blue, green, and black solid lines are theoretically obtained for the QMI, RHP, and BLP respectively, while the red and green dots are the respect experimental data for the QMI and the RHP.\label{fig:PhaseDiagram}}
\end{figure}

Because by varying $\omega_{0}$ we can effectively tune the temporal behavior of $\chi(t)$ \citep{Soare2014}, in Fig.~\ref{fig:PhaseDiagram} we investigate the non-Markovianity
vs the base frequency $\omega_{0}$. At a small $\omega_{0}$, e.g. $\omega_{0}=0.05$~MHz in Fig.~\ref{fig:PhaseDiagram}(b), the time evolution of the
norm of the off-diagonal term in the density matrix $f(t)$ manifests
a monotonic decay. The unidirectional information loss from the system
to the bath is a signature of a Markovian process. By increasing $\omega_{0}$,
$f(t)$ turns to be oscillatory as shown in Fig.~\ref{fig:PhaseDiagram}(c)-(d). To explore the underlying physical mechanism, we investigate the time dependence of $\chi(t)$ with respect to the change of the base frequency.
When $\omega_{0}$ is sufficiently small, $\chi(t)$ is quadratic with time. As $\omega_{0}$ increases, $\chi(t)$ becomes linearly dependent on time. However, if we further enlarges $\omega_{0}$, $\chi(t)$ will not be monotonically increasing with the time and thus results in temporal rise in $f(t)$.
Because the temporary rise in the off-diagonal term of the density matrix shows
the information backflow from the bath to the system, both these two processes
are not Markovian. Notice that the quantum dynamics in Fig.~\ref{fig:PhaseDiagram}(c) is Markovian until $t_f=5~\mu$s, as $f(t)$ decays monotonically in this period.
Hereafter, up to $t_f=5~\mu$s,
we explore the non-Markovianity by increasing $\omega_{0}$ as measured by the QMI, i.e., $\mathscr{\mathfrak{\mathcal{N}}}_{0}\left(\Lambda\right)=\smallint_{0,\frac{d}{dt}\mathcal{I}>0}^{t_f}\frac{d}{dt}\mathcal{I}\left(\rho_{\rm{sa}}\right)dt$
\citep{Luo2012}, which accumulates the rising part of the QMI, $\mathcal{I}=S\left(\rho_{\rm{s}}\right)+S\left(\rho_{\rm{a}}\right)-S\left(\rho_{\rm{sa}}\right)$ with $S(\rho)=-\textrm{Tr}(\rho\log_2\rho)$.
For a Drude-Lorentz spectral density with $F\left(\omega_{j}\right)=[2\lambda\gamma\omega_{0}\coth(\frac{\beta\omega_{j}}{2})/\omega_{j}(\omega_{j}^{2}+\gamma^{2})]^{1/2}$, $\alpha_{i}=\sqrt{\pi/2}$, reverse temperature $\beta=1/k_BT$, reorganization energy $\lambda$ and relaxation rate $\gamma$, the decoherence factor is
\begin{eqnarray}
\chi\left(t\right)=\lambda\gamma\omega_{0}\sum_{j=1}^{J}\frac{\coth\left(\frac{\beta\omega_{j}}{2}\right)}{\omega_{j}\left(\omega_{j}^{2}+\gamma^{2}\right)}\sin^{2}\frac{\omega_{j}t}{2}.
\end{eqnarray}
The reduced density matrices of the system and ancillary qubit are both $\rho_{\rm{s}}(t)=\rho_{\rm{a}}(t)=I/2$, while the nonvanishing eigen-values of the total density matrix are $\lambda_{\pm}=[1\pm f(t)]/2$.
The von Neumann entropy of the reduced density matrix of the system, ancilla and the composite system can be respectively given as
\begin{eqnarray}
S\left(\rho_{\rm{s}}\right)&=&S\left(\rho_{\rm{a}}\right)=1,\\
S\left(\rho_{\rm{sa}}\right)&=&-\frac{1+f\left(t\right)}{2}\log_{2}\frac{1+f\left(t\right)}{2}-\frac{1-f\left(t\right)}{2}\log_{2}\frac{1-f\left(t\right)}{2}.
\end{eqnarray}
The QMI and its derivative are explicitly given as
\begin{eqnarray}
\mathcal{I}(\rho_{\rm{sa}})&=&2+\frac{1+f(t)}{2}\log_{2}\frac{1+f(t)}{2}+\frac{1-f(t)}{2}
\log_{2}\frac{1-f(t)}{2},\\
\frac{d}{dt}\mathcal{I}\left(\rho_{\rm{sa}}\right)&=&-\chi^{\prime}\left(t\right)f\left(t\right)\log_{2}\frac{1+f\left(t\right)}{1-f\left(t\right)}.
\end{eqnarray}
Thus, the measure of non-Markovianity in terms of total correlation is
\begin{eqnarray}
\mathscr{\mathfrak{\mathcal{N}}}_{0}\left(\Lambda\right)=-\smallint_{0,\frac{d}{dt}\mathcal{I}>0}^{t_f}\chi^{\prime}\left(t\right)f\left(t\right)\log_{2}\frac{1+f\left(t\right)}{1-f\left(t\right)}dt.
\end{eqnarray}
In Fig.~\ref{fig:PhaseDiagram}(a), we show the transition
from the Markovianity to the non-Markovianity by tuning the base frequency. The transition occurring at $\omega_{0}=0.1$~MHz is consistent with the quantum dynamics in Fig.~\ref{fig:PhaseDiagram}(c), as $f(t)$ decays monotonically until $t_f=5~\mu$s. In Fig.~\ref{fig:PhaseDiagram}(a), we also compare the QMI with the BLP~\cite{Breuer2009-1} and the RHP~\cite{Rivas2010}. Interestingly, all three measures show a similar dependence on the $\omega_0$ and the phase transition in three cases accidentally occurs at $\omega_0=0.1$~MHz. These observations are consistent with the theoretical discoveries in Ref.~\cite{Jiang2013} that the three criteria for Markovianity may coincide in many cases, although they are different in general.

\bigskip
\subsection*{Quantum Fisher Information}

\begin{figure*}[tbh]
\centering
\begin{minipage}{\textwidth}
        \includegraphics[width=13cm]{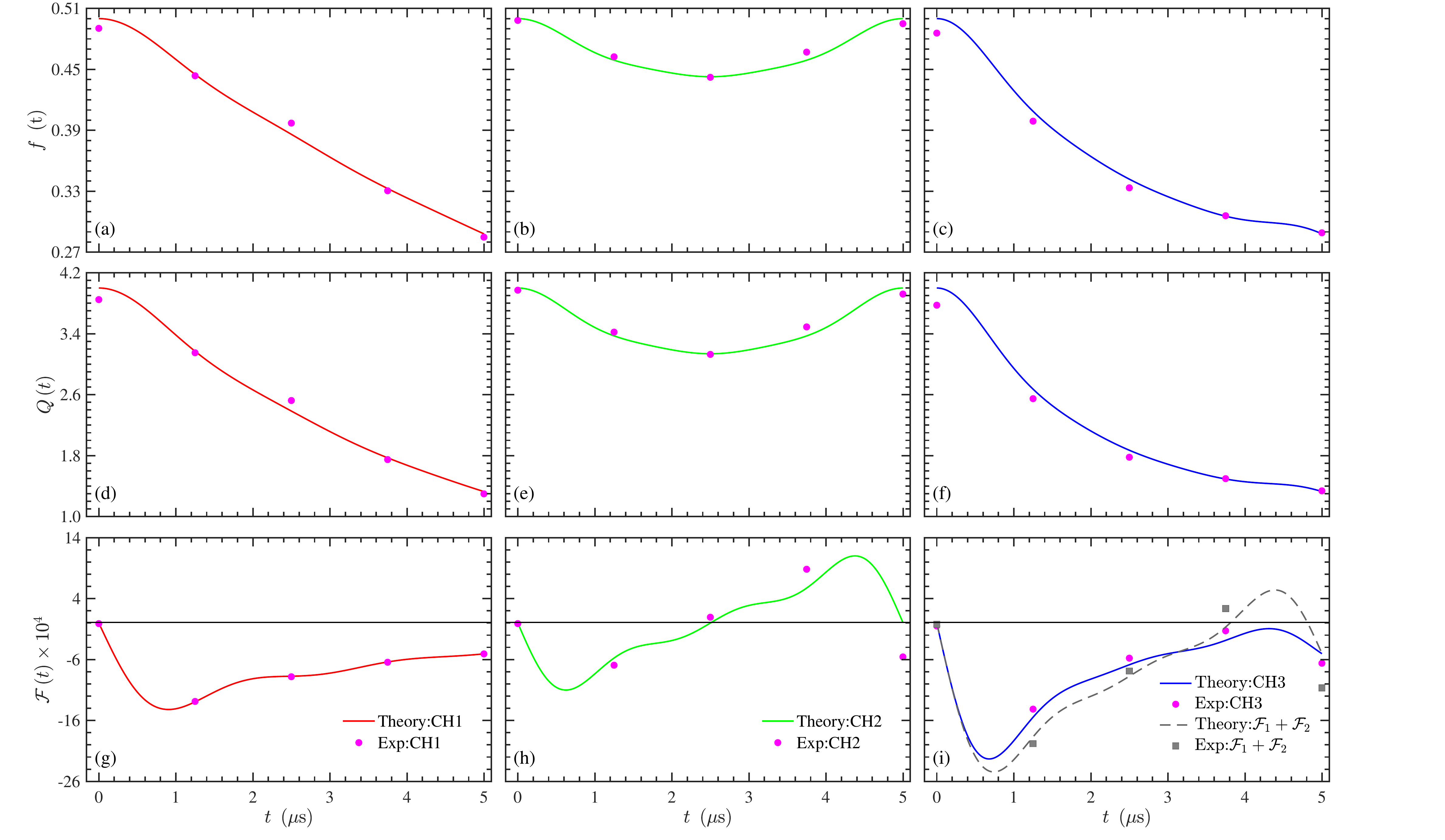}
        \caption{The QFI flows in different channels. Propagation of (a) off-diagonal term of the density matrix $f(t)$, (d) the QFI $Q(t)$, (g) the QFI flow $\mathcal{F}(t)$, when only CH1 is open. (b), (e), (h) are the corresponding results when only CH2 is open. (c), (f), (i) are the corresponding results when both of the two channels are open. }\label{fig:QFI}
    \end{minipage}
\end{figure*}

In addition to the QMI, the QFI flow was proposed to observe the information exchange between the system and bath in different channels \cite{Lu2010}.
Recently, Lu et al. engineered multiple dissipative channels of an open system in diamond and used QFI to quantify its non-Markovian dynamics \cite{Lu2020}. Here, we also use the QFI flow to analyze the local information exhange between the system and the bath in different channels. However, by virtue of bath-engineering technique, we can effectively engineer individual channels from Markovian to non-Markovian in the simulated open quantum dynamics. We further show that by mixing unitary dynamics, we could obtain non-Markovian open quantum dynamics \cite{Breuer2018}.
We consider a total system with two qubits, as shown in Fig.~\ref{fig:Scheme}(b), where the system qubit $^{13}$C is the first channel and the auxiliary qubit H is the second channel \cite{Lu2010}.
By applying noise to different qubits, we can realize the information flow of different channels.
When the noise is applied to the system and the auxiliary qubit, the corresponding channel 1 and channel 2 are opened respectively. When the noise is applied on both qubit, both channels are turned on.
We also assume that the initial state is the maximum-entangled state, i.e., $\vert\Phi_{\rm{sa}}(0)\rangle=(\vert00\rangle_{\rm{sa}}+\vert11\rangle_{\rm{sa}})/\sqrt{2}$.
The density matrix of the total system
at time $t$
can be diagonalized as
\begin{eqnarray}
\rho_{\rm{sa}}(t)=\sum_{i}P_{i}
{{\vert\psi_{i}\rangle\langle\psi_{i}\vert}},
\end{eqnarray}
where $\vert\psi_{i}\rangle$ is $i$th eigen-vector with eigen-value $P_{i}$.
For such a state $\rho_{\rm{sa}}$ and a generator $\hat{O}$, the QFI of a state $\rho_{\rm{sa}}(\theta)=\exp(-i\theta\hat{O})\rho_{\rm{sa}}\exp(i\theta\hat{O})$ with respect to a parameter $\theta$
can be given as \citep{Braunstein1994}
\begin{eqnarray}
Q=2\sum_{P_{i}+P_{j}\neq 0}\frac{(P_{i}-P_{j})^{2}}{P_{i}+P_{j}}
\vert\langle\psi_{i}\vert\hat{O}\vert\psi_{j}\rangle\vert^{2}.
\end{eqnarray}
When the noise is only applied on the system qubit,
i.e., ${H}=\frac{\omega_{\rm{s}}}{2}\sigma^{z}_{\rm{s}}+\frac{\omega_{\rm{a}}}{2}\sigma^{z}_{\rm{a}}+\beta_{\rm{s}}(t)\sigma^{z}_{\rm{s}}$,
the QFI and its flow with respect to the generator $\hat{O}=(\sigma^{z}_{\rm{s}}+\sigma^{z}_{\rm{a}})/2$ can be calculated as
\begin{eqnarray}
Q_{\rm{s}}&=&4\exp(-4\chi_{\rm{s}}), \\
\mathcal{F}_{\rm{s}}&=&-16\dot{\chi}_{\rm{s}}\exp(-4\chi_{\rm{s}}),
\end{eqnarray}
where the decoherence factor is $\chi_{\rm{s}}(t)=\frac{4}{2\pi}\int_{-\infty}^{\infty}\frac{d\omega}{\omega^{2}}
S_{\rm{ss}}(\omega)\sin^{2}(\frac{\omega t}{2})$.
When the noise is only applied on the auxiliary qubit,
i.e., ${H}=\frac{\omega_{\rm{s}}}{2}\sigma^{z}_{\rm{s}}+\frac{\omega_{\rm{a}}}{2}\sigma^{z}_{\rm{a}}+\beta_{\rm{a}}(t)\sigma^{z}_{\rm{a}}$,
the QFI and its flow are respectively
\begin{eqnarray}
Q_{\rm{a}}&=&4\exp(-4\chi_{\rm{a}}),\\
\mathcal{F}_{\rm{a}}&=&-16\dot{\chi}_{\rm{a}}\exp(-4\chi_{\rm{a}}),
\end{eqnarray}
where the decoherence factor is $\chi_{\rm{a}}(t)=\frac{4}{2\pi}\int_{-\infty}^{\infty}\frac{d\omega}{\omega^{2}}
S_{\rm{aa}}(\omega)\sin^{2}(\frac{\omega t}{2})$.
When both channels are turned on,
i.e., ${H}=\frac{\omega_{\rm{s}}}{2}\sigma^{z}_{\rm{s}}+\frac{\omega_{\rm{a}}}{2}\sigma^{z}_{\rm{a}}+\beta_{\rm{s}}(t)\sigma^{z}_{\rm{s}}+\beta_{\rm{a}}(t)\sigma^{z}_{\rm{a}}$,
the QFI and its flow are respectively
\begin{eqnarray}
Q_{\rm{sa}}&=&4\exp[-4(\chi_{\rm{s}}+\chi_{\rm{a}})], \\
\mathcal{F}_{\rm{sa}}&=&-16(\dot{\chi}_{\rm{s}}+\dot{\chi}_{\rm{a}})\exp[-4(\chi_{\rm{s}}+\chi_{\rm{a}})].
\end{eqnarray}
Notice that $\mathcal{F}_{\rm{sa}}\neq \mathcal{F}_{\rm{s}}+\mathcal{F}_{\rm{a}}$ implies that the overall QFI flow does not equal to the summation of the QFI flows from individual channels \cite{Lu2010}, as will be illustrated in Fig.~\ref{fig:QFI}(i).

By using the quantum simulation approach \cite{Wang2018,Zhang2020}, we again simulate the time evolution of density matrix $\rho(t)$ and compare it to that by the HEOM. The HEOM has been widely used in studying open quantum systems \cite{Ishizaki2009-2,Shi09,Liu14}, as it faithfully reproduces accurate open quantum dynamics. Nevertheless, the computational complexity of our quantum simulation method is significantly reduced as compared to that of the HEOM \cite{Wang2018,Zhang2020}.
By straightforward calculation, we obtain the QFI $Q(t)$ and then numerically fit the experimental data to attain the QFI flow $\mathcal{F}(t)=\partial_tQ(t)$. First of all, we turn on the noise of the system qubit. The off-diagonal term of the density matrix $f(t)$ decreases monotonically, cf. Fig.~\ref{fig:QFI}(a), and thus this channel is Markovian as the QFI flow is negative all the time in Fig.~\ref{fig:QFI}(g). Then, if we only switch on the noise of the auxiliary qubit, there are oscillations in $f(t)$, cf. Fig.~\ref{fig:QFI}(b), which suggests positive QFI flow in Fig.~\ref{fig:QFI}(h) when $f(t)$ is rising during the corresponding period. Finally, we examine the case when the noises are applied to both qubits. In Fig.~\ref{fig:QFI}(c,f), both of $f(t)$ and the QFI decrease monotonically along with the time. The process is Markovian since the QFI flow is always negative as shown in Fig.~\ref{fig:QFI}(i). However, the summation of the QFI flows over two individual channels can be positive and thus $\mathcal{F}_{\rm{sa}}\neq \mathcal{F}_{\rm{s}}+\mathcal{F}_{\rm{a}}$, because the total noise can exert influence on all channels \cite{Lu2010}. Therefore, we may safely arrive at the conclusion that it is inadequate to only use global measures such as entanglement and QMI to characterize the non-Markovianity of the open quantum dynamics. The local measures, e.g., the QFI flow, can subtly survey the non-Markovianity inside the open quantum system.

\bigskip
\section*{Discussion}

\begin{figure}
\includegraphics[bb=0 85 570 770,width=8.5cm]{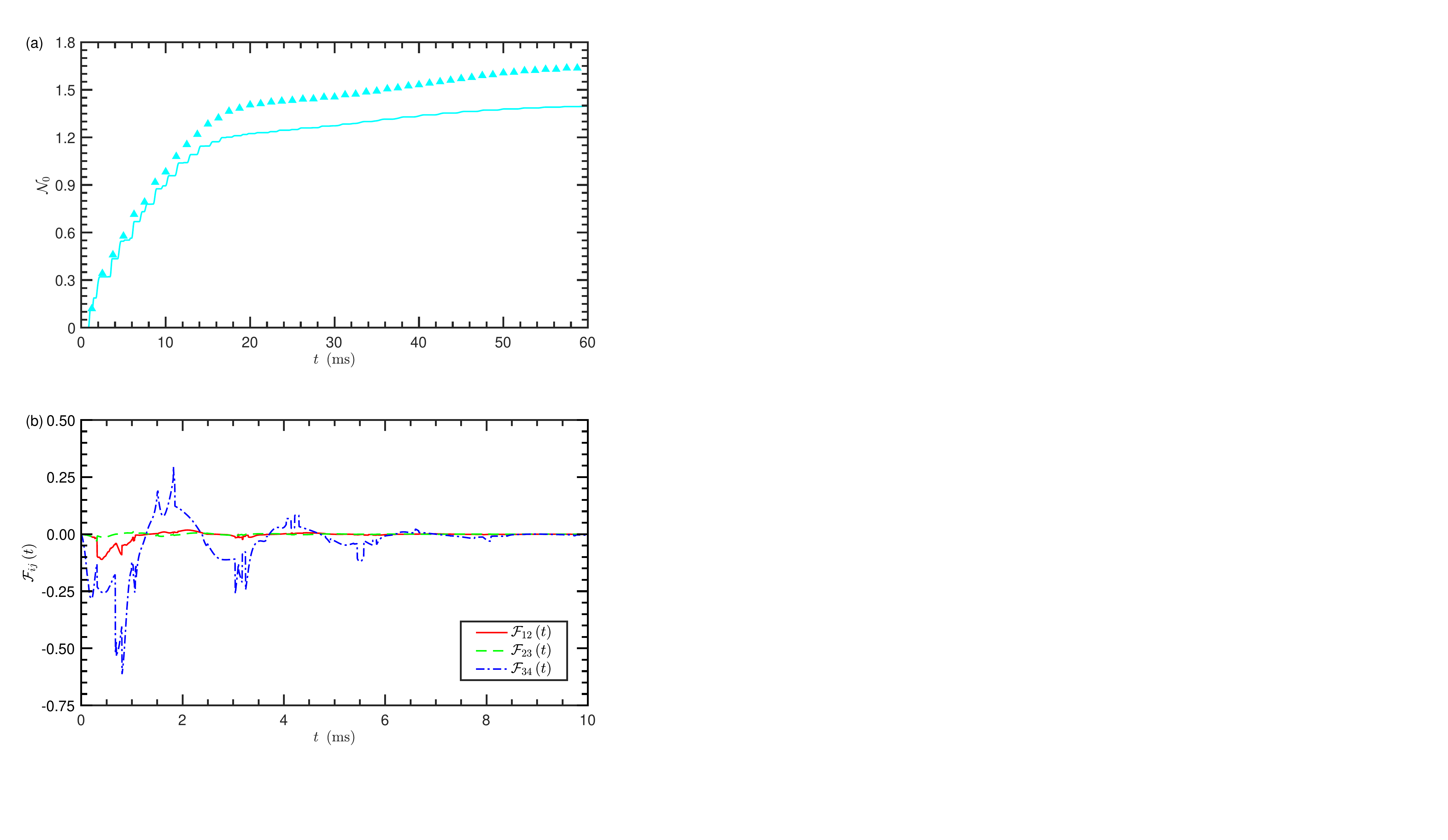}
\caption{Non-Markovianity in the photosynthetic energy transfer. (a) The QMI $\mathcal{N}_0$ and (b) the QFI flows from three dissipative channels $\mathcal{F}_{j,j+1}$ ($j=1,2,3$) in a generic two-qubit system by the quantum (symbol) and HEOM simulations (curve).\label{fig:generic}}
\end{figure}

In the above investigations, a specific model in which the noise Hamiltonian commutes with the system Hamiltonian is utilized in order to obtain the analytical results. However, in practice the system-bath interaction generally do not commutate with the system Hamiltonian. To show the broad applicability of our proposal, we use a generic model in Ref.~\cite{Ai2013}, where the environment assists efficient energy transfer in photosynthetic four-chromophore system. Since it has been shown in Refs.~\cite{Wang2018,Zhang2020} that the NMR experiments faithfully reproduce the exact results of the HEOM simulation, here we compare the time evolution of QFI flows $\mathcal{F}_{ij}(t)$ and QMI $\mathcal{N}_0$ by the quantum simulation and HEOM in Fig.~\ref{fig:generic}. In Fig.~\ref{fig:generic}(a), the QMI discontinuously rises and almost reaches the steady state around $t=18$~ms. The total QFI flow can be divided into the flows in different channels as $\mathcal{F}=\sum_{i,j}\mathcal{F}_{ij} =\sum_{i,j}\gamma_{ij}\mathcal{J}_{ij}$ \cite{Lu2010}, where $\mathcal{J}_{ij}=-\textrm{Tr}(\rho[L,A_{ij}]^\dagger[L,A_{ij}])$ with $\rho$ and $L$ being the density matrix and the symmetric logarithmic derivative respectively, $\gamma_{ij}$ is the decoherence rate associated with the jump operator $A_{ij}=\vert \varepsilon_i\rangle\langle\varepsilon_j\vert$ in the Lindblad-form quantum master equation with $\vert\varepsilon_i\rangle$ being $i$th eigen state of ${{H}_{\rm{S}}}$, i.e., ${{H}_{\rm{S}}}=\sum_i\varepsilon_i\vert\varepsilon_i\rangle\langle\varepsilon_i\vert$. Here, $\gamma_{ij}$ is approximated by the Redfield theory without the Markovian approximation \cite{Ishizaki2009-1}. The matrix elements of $L$ are given as $L_{ij}=2\langle \psi_i\vert\partial_\theta\rho\vert \psi_j\rangle/(p_i+p_j)$, where $\rho(t)=\sum_jp_j(t)\vert\psi_j(t)\rangle\langle\psi_j(t)\vert$.
In order to evaluate the QFI, we perform a unitary transformation $U(\theta)=\exp{[(\sigma^z_{\rm{s}}+\sigma^z_{\rm{a}})\theta]}$ on the initial state $(\vert00\rangle+\vert11\rangle)/\sqrt{2}$ before the open quantum dynamics \cite{Lu2010}, i.e., we adopt the same generator as in the second experiment. By varying $\theta$ with a small quantity $d\theta$, we could repeat the above procedure and obtain $\partial_\theta\rho$. Since $[{{H}_{\rm{S}},{H}_{\rm{SB}}}]\neq0$, there are generally two kinds of channels for the QFI, i.e., the dephasing and dissipative channels. In Fig.~\ref{fig:generic}(b), we only show the three dissipative channels responsible for the energy transfer between the nearest neighbours in the eigen basis, i.e., $\mathcal{F}_{j,j+1}$. Obviously, the QFI flows in the three channels are not equal. Because the coupling between $\vert1\rangle$ ($\vert3\rangle$) and $\vert2\rangle$ ($\vert4\rangle$) is larger than their site-energy difference, the energy explores coherent relaxation to transfer between them \cite{Ai2013}, and thus the $\mathcal{F}_{34}$ ($\mathcal{F}_{12}$) can be positive. Since $\vert00\rangle\equiv\vert1\rangle$ ($\sim\vert \varepsilon_4\rangle$) is the state (eigen state) with the highest site energy (eigen energy), $\mathcal{F}_{34}$ oscillates with the largest amplitude. On the contrary, since  $\vert11\rangle\equiv\vert4\rangle$ ($\sim\vert \varepsilon_1\rangle$) is the state (eigen state) with the lowest site energy (eigen energy), $\mathcal{F}_{12}$ oscillates with a smaller amplitude. However, due to the smaller coupling between $\vert2\rangle$ and $\vert3\rangle$ with respect to their large energy gap, the energy transfer between them is incoherent and thus results in a lower rate. Therefore, $\mathcal{F}_{23}$ is negative and its oscillation is negligibly small. Here, we show that although the QMI $\mathcal{N}_{0}$ indicates that the open quantum dynamics in the photosynthetic light-harvesting is not Markovian, the QFI flows subtly reveal the local characteristics for different channels.


In this paper, we apply a recently-developed experimental platform for the investigation on the non-Markovian open quantum dynamics. By tuning the base frequency, we can effectively change the open quantum dynamics from Markovianity to non-Markovianity. Furthermore, by turning on/off the noise on either qubit, we can observe the QFI flows from different channels. As a generalization, we apply these two measures on the investigations of the open quantum dynamics of photosynthetic energy transfer. In addition to the global measures such as the QMI, the local measure, i.e., the QFI flows, can show the local information backflow, which clearly demonstrate non-Markovianity from a local point of view.  Notably, we have experimentally realized mixing-induced non-Markovianity with unitary dynamics. Here, the random fields $\beta_i(t)$ essentially play the role as the ancillary system, which exchange information with the system, and thus the mixing-induced non-Markovianity emerges \cite{Breuer2018}. To conclude, we demonstrate that the combination of the QMI and QFI flow are fully capable of characterizing a multi-channel open quantum dynamics.

\section*{Methods}

\bigskip
\subsection*{Quantum Simulation}

In Refs.~\cite{Soare2014,Zhang2020}, both the pure-dephasing and energy relaxation form of system-bath interaction have been experimentally realized for various of spectral densities.

In order to simulate the dynamics of open quantum systems in NMR,
we apply a time-dependent
local magnetic field $\beta_{i}=\alpha_{i}\sum_{j=1}^{J}F_{i}(j)\sin(\omega_{j}t+\psi^{(i)}_{j})$
to the state $\vert i\rangle$, i.e.,
\begin{eqnarray}
{{H_{\mathrm{QS}}}}={{H}}_{\mathrm{S}}+{{H}}_{\mathrm{N}},
\end{eqnarray}
where
${{H}_{\rm{S}}}$ and
${{H}_{\rm{N}}}=\sum_i\beta_i\sigma_i^{z}$ correspond to the system Hamiltonian and the noise Hamiltonian, respectively.
In the simulation, we initially
prepare a large number of ensemble in the same initial state, which
afterwards are subject to different Hamiltonians ${{{H_{\mathrm{QS}}}}}$'s determined
by $\{ \psi_{j}^{(i)},j=1,\cdots,J\} $. By averaging over the ensemble, we
obtain the exact time evolution of the density matrix of the open quantum system predicted by the hierarchical equation of motion (HEOM), if we assume the system Hamiltonians in the open quantum dynamics and quantum simulation are the same, and the spectral density and power spectral density are equal \citep{Wang2018,Zhang2020}. Physical speaking, the coherent and incoherent dynamics are determined by the system Hamiltonian and spectral density, respectively. As long as these two quantities are the same, the open quantum dynamics can be faithfully reproduced by the quantum simulation. Finally, the theoretical Hamiltonian ${{H_{\mathrm{QS}}}}$ is decomposed into a series of experimentally feasible pulse sequences by the GRAPE algorithm. In this way, we can apply our quantum simulation approach for efficient simulating the exact open quantum dynamics.

\bigskip\noindent
\textbf{DATA AVAILABILITY}\\
The data that support the findings of this study are available from the
corresponding author QA upon reasonable request.

\bigskip\noindent
\textbf{ACKNOWLEDGEMENTS}\\
We thank stimulating discussions with X.-M. Lu, B.-X. Wang and J.-W. Wen.
This work was supported by the National Basic Research
Program of China under Grant No.~2017YFA0303704, the National Natural
Science Foundation of China under Grant Nos.~61727801, 11774197, 11474181,
11674033, 11505007, 11474026, and Beijing Natural Science Foundation
under Grant No.~1202017. N.-N. Zhang was partially supported by Chongqing University of Posts and Telecommunications under Grant No.~A2021-263.

\bigskip\noindent
\textbf{COMPETING INTERESTS}\\
The authors declare no competing interests.

\bigskip\noindent
\textbf{AUTHOR CONTRIBUTIONS}\\
X.Y.C., N.N.Z. and W.T.H. contributed equally to this work.
All work was carried out under the supervision of G.L.L., Q.A., and F.G.D.
X.Y.C. and X.Y.K. performed the experiments.
N.N.Z., W.T.H., M.J.T., X.Y.C. and Q.A. analysed the experimental data.
N.N.Z., W.T.H., M.J.T. and X.Y.C. wrote the HEOM simulation code
and performed the numerical simulation.
All authors contributed to writing the manuscript.



\noindent\textbf{Supplementary information}
See Supplementary Information for the details about theoretical deduction, numerical simulation, and complete experimental data.


\end{spacing}
\end{document}